# Photon counting with photon number resolution through superconducting nanowires coupled to a multi-channel TDC in FPGA

N. Lusardi,[1] J.W.N. Los,[2] R.B.M. Gourgues,[2] G. Bulgarini[2] and A. Geraci[1]

[1]*Politecnico di Milano, Department of Electronics-DEIB, via C. Golgi 40, Milan, 20133, Italy*
[2]*Single Quantum, Lorentzweg 1, 2628CJ, Delft, Netherlands*

The paper presents a system for measuring photon statistics and photon timing in the few-photon regime down to the single-photon level. The measurement system is based on superconducting nanowire single photon detectors and a time-to-digital converter implemented into a programmable device. The combination of these devices gives high performance to the system in terms of resolution and adaptability to the actual experimental conditions. As case of application, we present the measurement of photon statistics for coherent light states. In this measurement, we make use of 8[th] order single photon correlations to reconstruct with high fidelity the statistics of a coherent state with average photon number up to 4. The processing is performed by means of a Time-to-Digital Converter (TDC) architecture that also hosts an Asynchronous-Correlated-Digital-Counter (ACDC) implemented in a Field Programmable Gate Array (FPGA) device and specifically designed for performance optimization in multi-channel usage.

## I. INTRODUCTION

Extremely sensitive light sensing can be performed by observing the transition of a section of a current-biased superconducting nanowire, which absorbed a photon, from the superconducting state to the normal resistive state. Devices based on this principle are called Superconducting nanowire single photon detectors (SNSPDs) and find application in several areas of quantum information technology [1] for their ability of detecting single photons with near unity probability [2] in a large wavelength range from UV to infrared. SNSPDs have been used to measure single-photon emission from a wide variety of light sources as, for example, individual dopants in carbon nanotubes [3], color centers in silicon carbide [4] and semiconductor quantum dots [5]. In addition to high photon detection efficiency, SNSPDs provide several advantages in comparison to other single-photon sensitive devices: extremely low timing jitter [6], the absence of afterpulsing and the low dark count rate. Such performances make them the ideal detector for applications where efficient detection of weak signals with high time resolution is required. Examples include: high-resolution light detection and ranging (Lidar) [7], oxygen singlet detection [8], optical time domain reflectometry in telecommunication networks [9] and deep-space optical communication [10].

In this work, timing measurements of detected pulses are performed by means of a Time-to-Digital Converter (TDC) implemented in a Field Programmable Gate Array (FPGA) device. From the side of pure economic and design convenience, the realization in FPGA is by far preferable due to hugely lower development costs and extremely higher flexibility of the implemented architecture thanks to the programmable resources of the device. Main feature of the developed architecture is to be specifically addressed to multi-channel applications with the use of minimum resources and power supply overhead.

## II. DESCRIPTION OF THE DETECTOR

The detector has the shape of a meandering nanowire of NbTiN with 100 nm width and 50% fill factor, covering a circular area of diameter 16 µm. The superconductor is positioned on top of a resonant cavity formed by a silicon-oxide layer of 135 nm and gold in order to enhance the photon absorption probability at ~800 nm [11]. This wavelength range is particularly important for quantum experiments since both state-of-the-art solid-state single photon sources and those based on photon down-conversion provide photon emission in this wavelength range. The detector is operated at a temperature of ~2.5 K and is coupled to a single-mode optical fiber. Figure 1 shows the calibration of one superconducting nanowire detector used in this experiments. The system detection efficiency measured at 878 nm is plotted as a function of the bias current applied to the superconducting sensor.

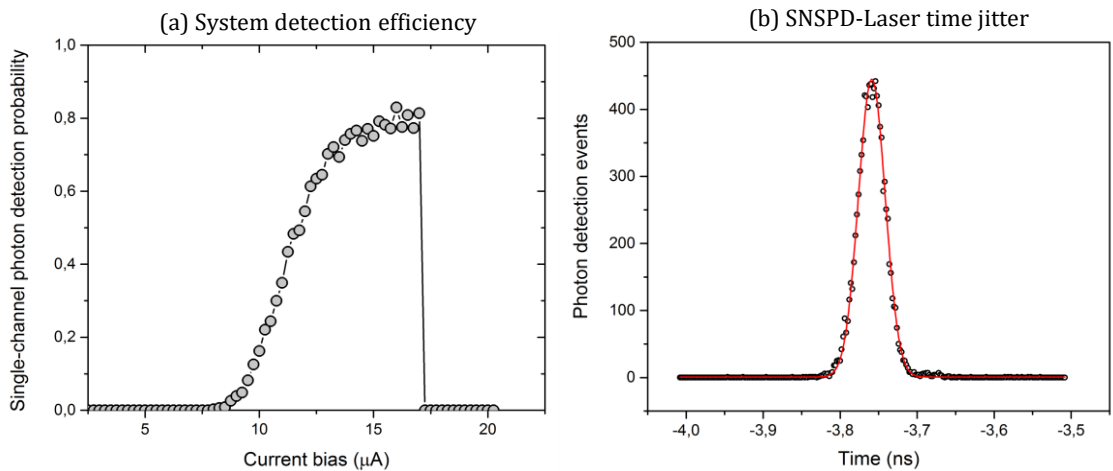

FIG. 1. (a) System detection efficiency measured for a single channel as a function of the bias current. The probability of detecting photons is increased as the current bias approaches the device critical current at 17 µA. (b) Histogram of the Laser-CH2 interval measurements performed with oscilloscope in order to estimate the contribution to jitter of device and amplifiers. Timing jitter is 19.84 ps r.m.s. In details, the contribution of the electrical signal from the laser is 6 ps r.m.s., the oscilloscope contribution is <2 ps r.m.s.

The device reaches at saturation a system detection efficiency of 80.7%. This efficiency represents the ratio of detected photon rate with respect to the photon rate at the cryostat input. It therefore includes optical losses at the fiber connector located at the cryostat input and it also includes the transmission losses of the optical fiber that connects the superconducting detectors from inside the cryostat to the cryostat input. Figure 1b displays the measured time distance of a laser trigger from a mode-locked Ti:Sapphire laser. The distribution of detection pulses shows a timing jitter ($\sigma_{jitter}^{SC}$) of 19.84 ps r.m.s., which corresponds to 46.62 ps FWHM for a Gaussian fit. The measurement in Figure 1b is acquired with a LeCroy Waverunner 640Zi oscilloscope. The electrical jitter of the laser trigger signal ($\sigma_{LASER}$) has been measured separately and resulted in a contribution of less than 6 ps r.m.s. corresponding to 14 ps FWHM. The contribution of the oscilloscope ($\sigma_{SCOPE}$<2 ps r.m.s. that is <4.7 ps FWHM) can be neglected. The jitter contribution of the SNSPD ($\sigma_{SNSPD}$) is below 8 ps r.m.s. and of the amplification stage ($\sigma_{AMPLI}$) is less than 14 ps r.m.s., i.e.

$$\sigma_{jitter}^2 = \sigma_{SNSPD}^2 + \sigma_{AMPLI}^2 + \sigma_{LASER}^2 + \sigma_{SCOPE}^2 \cong \sigma_{SNSPD}^2 + \sigma_{AMPLI}^2 + \sigma_{LASER}^2 \qquad (1)$$

## III. EXPERIMENTAL SETUP

The timing circuit performing the measurement is a resource-saving 8-channels Time-to-Digital Converter (TDC) [12] and an Asynchronous-Correlated-Digital-Counter (ACDC) implemented in the Programmable Logic section (PL) of a System-on-Chip (SoC) programmable device, i.e. a Xilinx ZYNQ-7020, that integrates a 28 nm Xilinx's FPGA within the Programmable Logic (PL) section and an ARM-based processor in the Programmable Software section (PS). The setup is composed of five main parts (Fig.2), which are a detection section, an analog stage, a group of comparators for digitizing the output of the analog stage, the SoC that includes TDC and ACDC and a read-out software running on a host PC.

The signal at the output of each detector has decreasing exponential shape with peak amplitude of about 0.5 mV and nominal decay time constant equal to 15 ns (Fig.5d). The analog stage is composed of 8 (one per channel) AC-coupled pass-band and low-noise double-stage amplifiers, with gain 50 dB over the bandwidth 10MHz-1GHz adding timing jitter to the intrinsic jitter of the single SNSPD that is less than 8 ps r.m.s. [6]. Each output of the 8 analog amplifiers is sampled by an ultra-fast and low-jitter comparator (Analog Devices ADCMP607BCPZ), whose output is a logic level in LVDS format. The timing jitter introduced by each comparator ($\sigma_{CMP}$) is below 7 ps r.m.s.. As stated above, The multi-channel TDC and the ACDC are implemented in a commercial 28 nm programmable device (Xilinx ZYNQ-7020). Main features of the TDC architecture are the low power consumption (430mW), the very modest expense of resources necessary for implementation (30 % of the PL section), and the resolution guaranteed below 15 ps r.m.s. per channel ($\sigma_{TDC}$). The scale-range of the TDC is limited to 640 ns because the maximum time interval under measurement is the time skew between two different channels, which is lower than few tens of ns.

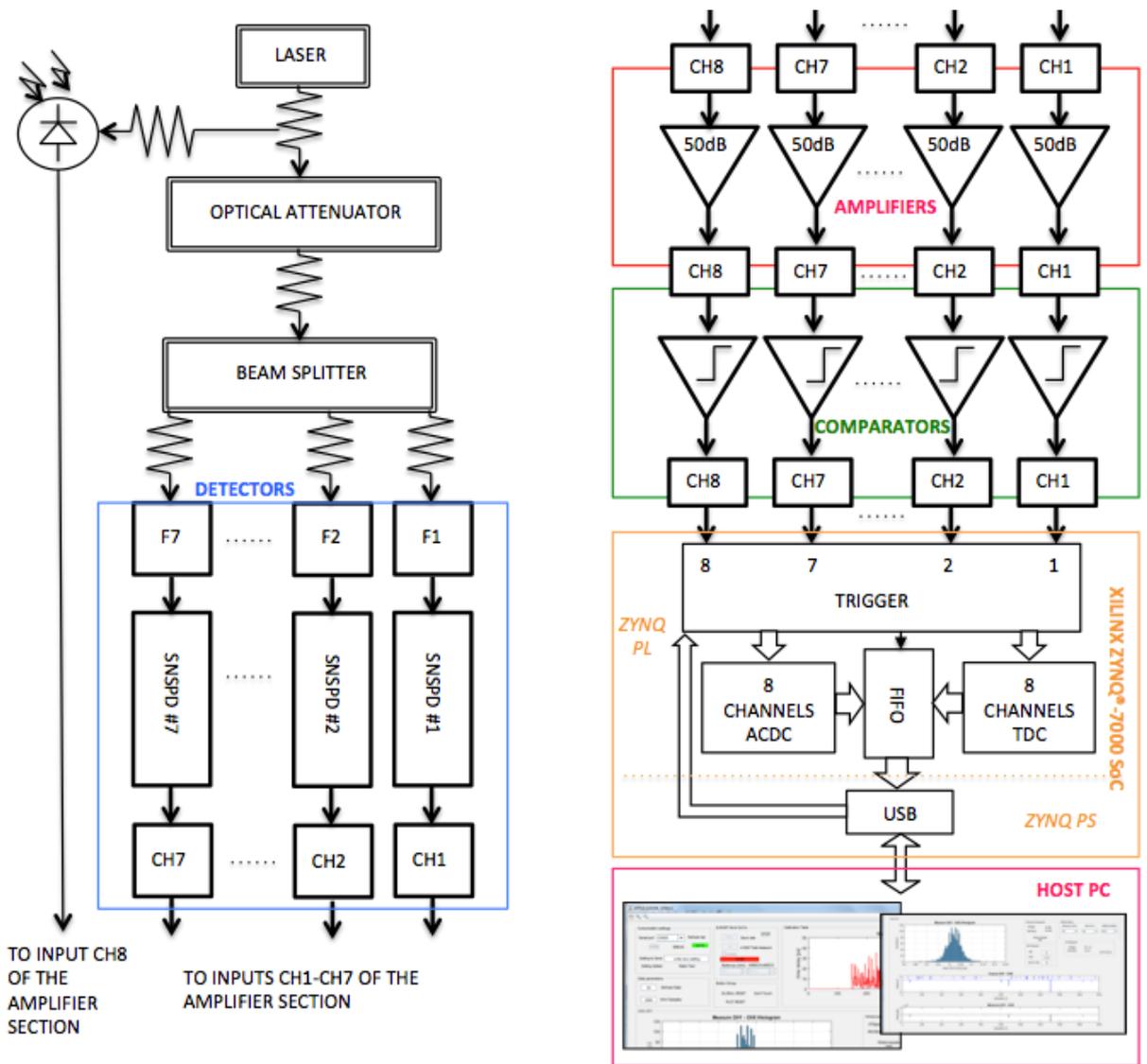
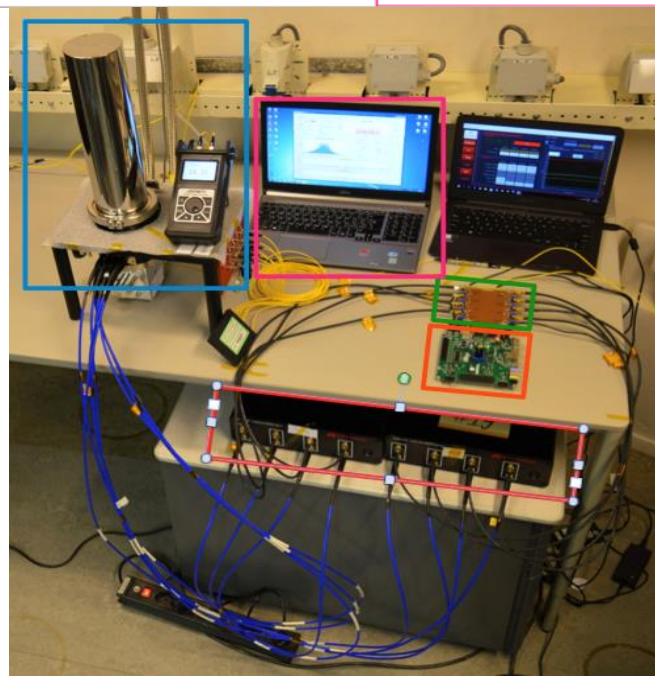

FIG. 2. Schematic description of the measurement setup and, below, picture of the setup with details marked according to scheme above.

The ACDC has 8 channels and detects the presence into the channel of an event for each single acquisition. The trigger maximum time resolution is 2.5 ns, which is significantly smaller than the maximum rate of detection in our experiment due to the temporal distances of the lasers pulses (i.e. 12.6 ns).

One channel is set as trigger of the acquisition process and gives the start for the photon counting on the other channels for a programmable time window (Fig.3). The SoC device sends time and counting data to the host PC via USB. The user via software sets the specific channel dedicated for triggering and the time duration of the measurement.

By means of the read-out software hosted on a standard computer, the user can set which channel is the "start" of the acquisition and the duration by a programmable time window.

In order to minimize the total jitter, it is worth using an acquisition time that lasts at least the repetition period of the laser pulse. Consequently, the procedure involves the acquisition of the 7 channels after a laser pulse triggers the one selected.

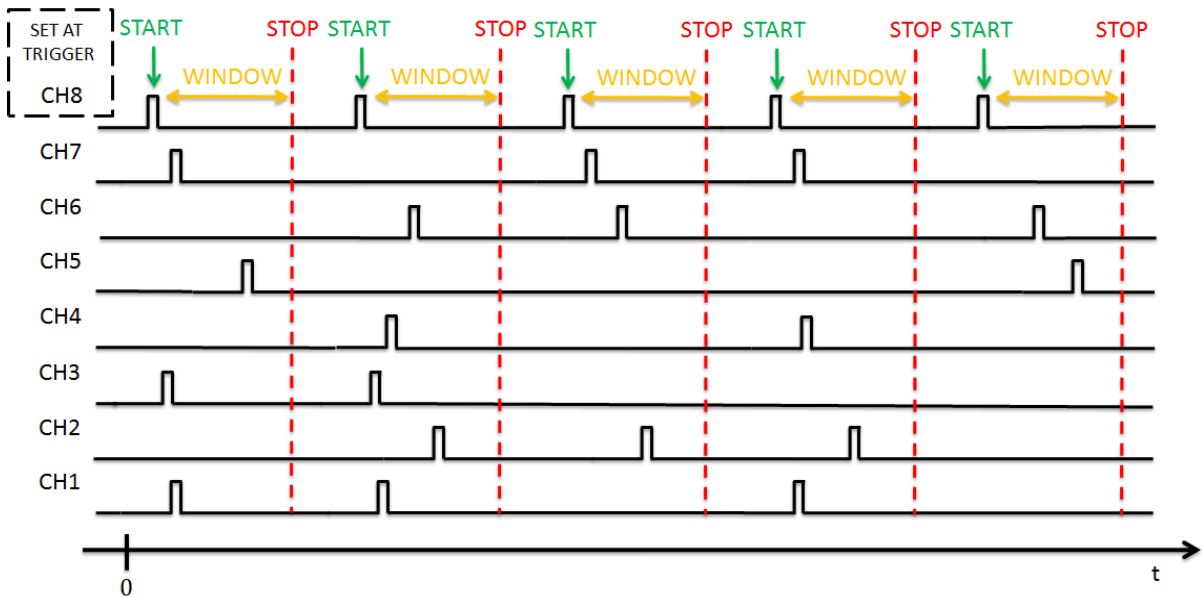

FIG. 3. Time diagram of an acquisition. In correspondence to each START event on the channel selected as trigger (e.g. channel CH8), the time intervals between START and the events detected by the ACDC into the acceptance WINDOW are calculated by the TDC.

## IV. DETECTION OF THE PHOTON TIMING IN A WEAK OPTICAL PULSE

As first experiment, we propose the calculation of the statistic of the times of arrival of the photons from a weak laser pulse with the purpose of measuring the time resolution of our complete system. High time resolution at the single photon level is of fundamental importance for analyzing fast photon emission dynamics at low illumination fluxes or in the presence of quantum emitters. An application example is the measurement with high contrast two-photon quantum interference [5]. In lidar and optical time-domain reflectometry, distances are determined by the time-of-flight measurement of optical signals reflected by the target. In this case, the accuracy of the distance measurement is directly determined by the time resolution of the measurement.

The laser source used in our experiment is a Ti:Sapphire laser with 76 MHz repetition rate, 6 ps pulse width, and 750 nm wavelength. Referring to the layout shown in Figure 2, the TDC measures the temporal distances of the events on the channel from the reference trigger on channel 8, which is given by the internal fast photodiode of the laser. The time interval and its temporal statistics from two generic channel can be shown on the host PC via the read-out software.

In order to identify the contributions to the jitter of the different components, first, we have bypassed the detector and the analog stage by entering directly in the comparators with the signals of a function generator. In order to be faithful to experimental conditions, the signals are generated in the actual range of frequency rates, i.e. from 50 kHz to 1.8 MHz, and with the same temporal shape as the output of the superconducting nanowire detectors. Table I shows a synoptic view of the resolutions of all channels measured taking channel 8 (LASER) as reference. In all other possible selections of the trigger channel and considering every combination of couples of channels, the cascade of comparators and TDC maintains the couple of channels resolution below 25 ps r.m.s. that is < 17 ps r.m.s. for each single channel ($\sigma_{TDC}^2 + \sigma_{CMP}^2$).

TABLE I. Standard deviation $\sigma_{ij}$[ps] of the measurement of the time interval $T_{ij}$[ns] between START and STOP events on channels CH$i$-CH$j$.

|  | CH1-CH8 | CH2-CH8 | CH3-CH8 | CH4-CH8 | CH5-CH8 | CH6-CH8 | CH7-CH8 |
|---|---|---|---|---|---|---|---|
| $T_{ij}$ [ns] | 1.81 | 2.67 | 2.83 | 3.36 | 3.08 | 2.70 | 2.80 |
| $\sigma_{ij}$ [ps] | 23.1 | 22.2 | 23.1 | 23.2 | 24.0 | 23.6 | 22.7 |

By measuring the complete detection and timing channel, i.e. including the superconducting nanowire detector and the analog circuitry in the measurement path, we observe that the minimum achievable resolution approximately doubles. Table II sums up the resolutions of all complete channels measured taking channel 8 as reference. The block of comparator and TDC guarantees the timing resolution between couples of channels below 58 ps r.m.s. for input rates varying from tents of kHz to a few MHz and for all possible START-STOP combinations, using any channel as trigger channel. It should be considered that the variance of the measurement $\sigma_{ij}^2$ is the sum of variances of channels $i$ and $j$ respectively, which means that the resolution of a single channel can be estimated to be $\sigma_{ij}/\sqrt{2}$.

TABLE II. Standard deviation $\sigma_{ij}$[ps] of the measurement of the time interval $T_{ij}$[ns] between START and STOP events on channels CH$i$-CH$j$.

|  | CH1-CH8 | CH2-CH8 | CH3-CH8 | CH4-CH8 | CH5-CH8 | CH6-CH8 | CH7-CH8 |
|---|---|---|---|---|---|---|---|
| $T_{ij}$ [ns] | 4.21 | 4.74 | 6.60 | 5.92 | 5.53 | 5.57 | 6.26 |
| $\sigma_{ij}$ [ps] | 51.5 | 46.7 | 51.7 | 47.6 | 56.9 | 57.7 | 55.5 |

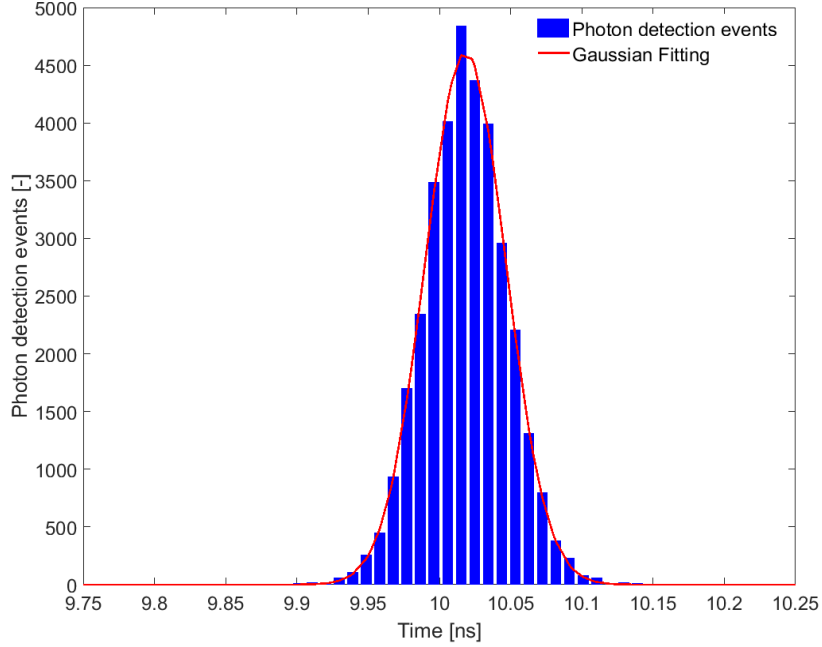

FIG. 4. Histogram of measurement of the time interval limited by START (channel 2)-STOP (channel 5) events on different SNSPDs. The resolution is 28.33 ps r.m.s.

Next, we consider the baseline fluctuation ($\sigma_{BASE}$ estimated <40 ps r.m.s.) due to a stochastic rate of arrival of the events and its impact on the resolution. The baseline fluctuation occurs because the photon detector transforms the deterministic frequency of 76 MHz of the laser into an equivalent lower stochastic rate. We restore the baseline by applying a high-pass filter at 1 MHz at the input of the TDC. Hence, the resolution is reduced to below 30 ps r.m.s. between couples of channels (Fig.4), which means 22 ps r.m.s. per channel ($\sigma_{CH}^2$). The improvement respect to the 50 ps of Tab. II is considerable. Moreover, this result is consistent with the best achievable value due to the nominal jitter of the individual components constituting the system that is calculated to be around 23.1 ps r.m.s.

$$\sigma_{ij}^2 = \sigma_{CHi}^2 + \sigma_{CHj}^2 + \sigma_{BASE}^2 \qquad (2)$$

$$\sigma_{CH}^2 = \sigma_{SNSPD}^2 + \sigma_{AMPLI}^2 + \sigma_{CMP}^2 + \sigma_{TDC}^2 \cong 8^2 + 14^2 + 7^2 + 15^2 \cong 23.1^2 \qquad (3)$$

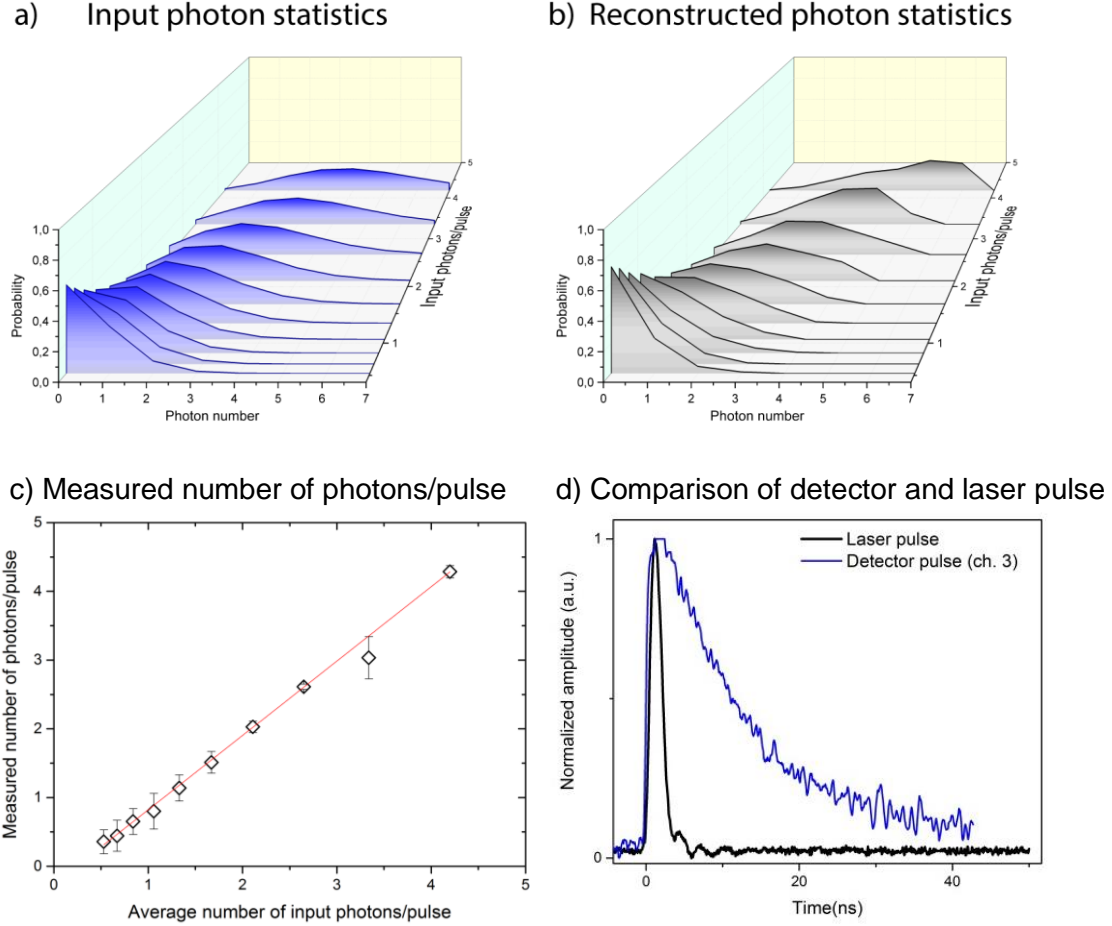

FIG. 5. (a) Input photon statistics at the beam splitter over several different measurements. The photon number probability for attenuated laser pulses follows a pure Poissonian statistics. We limit our investigation up to an average photon number of ~4 photons/pulse in order to have negligible probability of having a 7 photons per pulse, that is the number of detectors used in the experiment. (b) Photon statistics reconstructed from the data after normalization of the detection events by the finite detection probability. (c) Measured number of photons/pulse as compare to the input photon number. The data is fitted with a linear fit with $R^2$ exceeding 0.99. (d) Comparison of the voltage output pulse of the detector and the laser pulse used in the experiment.

## V. DETECTION OF THE PHOTON STATISTICS IN A WEAK OPTICAL PULSE

We now show a measurement that employs simultaneously several SNSPDs devices connected to the TDC/ACDC unit as displayed in Figure 2. Thanks to the ACDC part, we measure the number of photons in weak optical pulses by measuring the photon statistics. Photon number resolution in the few-photon regime is a crucial feature for the development of quantum information technologies like quantum cryptography. Since secret keys are exchanged by means of single photons or photon pairs, an eavesdropper could get information on the secret key if excess photons are transmitted through the link due to imperfect photon sources. In order to maintain quantum protection against potential eavesdroppers, it is necessary to employ a tool detecting the number of photons used over the communication link and the corresponding photon statistics.

Since the absorption of more than one photon simultaneously in a SNSPD gives the same electrical output as if one photon is absorbed, a single SNSPD cannot distinguish the precise number of photon hitting the detector. In our experiments, we overcome this limitation by spatially multiplexing the optical signal onto 7 different SNSPDs installed in a cryostat based on a Gifford-McMahon closed-cycle cryo-cooler and making use of the multi-channel TDC in FPGA architecture. By measuring on the TDC the photon arrival time on the 7 detectors simultaneously as compared to the laser trigger, we are able to determine the number of detection events per each laser clock. Through a calibration of the photon detection probability, we are able to reconstruct the input photon statistics from the acquired histogram of coincident detection events that is build up to the $8^{th}$ order coincidence.

Other ways to measure the photon number in a weak optical pulse are to use parallel detectors to output electrical pulses of different height as a function of the number of detectors transitioning to the normal state [13-16]. This technique requires the detectors to be operated at a much lower current than the critical current, thereby moving away from the optimum efficiency point. Alternatively, one can apply schemes for time multiplexing of the photons [17,18]. An incident pulse is divided in time into N weaker pulses by a series of fiber delay lines. If sufficiently long fibers are used, in order to delay the photons for a value longer than the detector dead time, this architecture enables for using a limited number of photon detectors. However, this implementation is not easily scalable for high photon number and it is not suitable for high repetition rates, i.e. high photon fluxes. Our approach makes use of a scalable TDC architecture that is connected to several high efficiency photon detectors. Our architecture can be readily scaled up to 32 detection channels because of the used FPGA architecture. The device can make full use of the superconducting nanowire performance of photon detection and time resolution.

The optical signal is obtained by a laser diode with wavelength of 878 nm, electrically pulsed at a repetition rate of 100 kHz, with a pulse width of 2 ns. The width of the optical pulse is approximately one order of magnitude smaller than the device dead time as shown in Figure 5d. The dead time is measured as the time constant of the decay of the voltage output pulse, which is due to the recovery of superconductivity in the detector after a detection event. The width of the pulse is kept much shorter than the detector dead time to avoid consecutive detection events from a single detector during one measurement window. The average laser power is measured with a calibrated optical power meter and then attenuated by a tunable optical attenuator for achieving an average of a few photons per pulse. The Poissonian distribution of photons within the weak optical pulses arriving at the beam splitter is shown in Figure 5a from an average photon number of 0.5 to 4.2. The number of sensors used in the experiments sets a limit to the average photon number that the device is able to measure.

For the measurement, the laser is connected to an eight-port fiber beam splitter that multiplexes the input to 7 detectors. One port of the splitter is not used since the electrical signal corresponding to the laser trigger is directly connected to channel 8 of the TDC. We note that it is essential that one TDC channels acquires the laser trigger to open a detection window. The events where no detectors click, after the laser trigger has been acquired, will be assigned to the 0-photon probability. Each detector, as well as the transmission of each fiber, has been separately calibrated. The efficiency of each detector and the transmission of the fiber splitter have been used, for each input power used in Figure 4a, to calculate the system response and thus to reconstruct the photon statistics of the optical pulse arriving at the fiber splitter. The reconstructed photon statistic for each input power is shown in Figure 4b after the acquisition of 10000 photon pulses. The measured photon number inferred by our device as compared to the photon number measured with optical attenuator and calibrated optical power meter is shown in Figure 4c. The results are shown with a linear fit with $R^2>0.99$. Additionally, we observe that the relative error in the determination of the photon number is decreased with increasing number of photons per pulse, since the signal-to-noise ratio (i.e., photon count rate versus dark count rate). The average dark count rate per channel in this experiment is between 30 and 50 Hz. The results demonstrate that using superconducting nanowire detectors coupled to a multi-channel TDC we can perform up to 8-fold photon coincidence measurements and measure with the ACDC the photon statistics of coherent states of light arriving at the beam splitter.

## VI. CONCLUSIONS

A setup for measuring photon statistics and photon timing in the few-photon regime down to the single-photon level has been presented. The system is based on SNSPDs and a TDC architecture hosting an ACDC implemented in a FPGA device. Main target applications are several areas of quantum information technology for the feature of detecting single photons at high efficiency in a wide wavelength range from UV to infrared, with extremely low timing jitter, absence of afterpulsing and low dark count rate.

The processing of detected pulses in the FPGA device has the advantage of enabling flexible implementation of different architectures thanks to the programmable resources, which in our system have been focused towards synthetizing a re-configurable multi-channel structure with minimum resource and power supply overhead. It should be considered that the option of using an oscilloscope for START-STOP measurements requires the presence of the laser reference for avoid complex and not always possible triggering methods. In this case the ACDC component is not present and consequently the function of calculating the statistics of incoming photons is not available. Moreover, with an oscilloscope the number of channels measuring in parallel is limited using the FPGA is possible extend the number of channel keeping the resolution almost constant.

Referred to the single channel, the complete measurement system has resolution below 22ps r.m.s., power consumption of 2.5mW, and employing only 2.7% on the selected FPGA device ZYNQ7020. The trigger resolution is equal to 2.5 ns. The measurement system presented in this article could be promptly scaled up to perform timing measurements over 32 channels simultaneously, calculating statistics and detecting time correlations.

**ACKNOWLEDGMENTS**

J.W.N. Los, R.B.M. Gourgues, and G. Bulgarini are thankful to S. Dobrovolskiy for the fabrication of superconducting photon detectors. The work of J.W.N. Los, R.B.M. Gourgues, and G. Bulgarini is supported by the European Commission through the H2020 programme through SUPERTWIN project, id. 686731. R.B.M. Gourgues acknowledges support by the European Commission via the Marie-Sklodowska Curie action Phonsi (H2020-MSCA-ITN-642656).